\documentclass[fleqn]{aa}
\usepackage{amsmath}
\usepackage{amssymb,amsfonts}
\usepackage[mathscr]{eucal}
\usepackage{astron}
\usepackage{epsfig}

\begin{document}   
\thesaurus{(03.13.2;12.03.1;02.16.2) }
\authorrunning 
\titlerunning

\title{Destriping of Polarized Data in a CMB Mission with a Circular Scanning Strategy}

\author{B.~Revenu \inst{1} \and A.~Kim \inst{1} \and R.~Ansari \inst{2} \and F.~Couchot \inst{2} \and
  J.~Delabrouille \inst{1}  \and J.~Kaplan \inst{1}}

\offprints{B. Revenu (revenu@cdf.in2p3.fr)}

\institute{
Physique Corpusculaire et Cosmologie, Coll{\`e}ge de France, 11 Place Marcelin Berthelot, 75231 Paris Cedex 05, France
\and
Laboratoire de l'Acc{\'e}l{\'e}rateur Lin{\'e}aire, IN2P3 CNRS, Universit{\'e} Paris Sud, 91405 Orsay, France
}

\date{Received; accepted }
\maketitle

\markboth{Revenu et al.}{Destriping of CMB polarized data}

\begin{abstract}
  
  A major problem in Cosmic Microwave Background (CMB) anisotropy
  mapping, especially in a total-power mode, is the presence of
  low-frequency noise in the data streams. If unproperly processed,
  such low-frequency noise leads to striping in the maps. To deal with
  this problem, solutions have already been found for mapping the CMB
  temperature fluctuations but no solution has yet been proposed for
  the measurement of CMB polarization. Complications arise due to the
  scan-dependent orientation of the measured polarization. In this
  paper, we investigate a method for building temperature and
  polarization maps with minimal striping effects in the case
  of a circular scanning strategy mission such as the {\sc
  Planck} mission.

\keywords{methods: data analysis; cosmology: cosmic microwave background;
polarization}
\end{abstract}
\bibliographystyle{astron}

\section{Introduction}

Theoretical studies of the CMB have shown that the accurate
measurement of the CMB anisotropy spectrum $C^T_\ell$ with future
space missions such as {\sc Planck} will allow for tests of cosmological
scenarios and the determination of cosmological parameters with
unprecedented accuracy. Nevertheless, some near degeneracies between
sets of cosmological parameters yield very similar CMB temperature
anisotropy spectra.  The measurement of the CMB polarization and the computation of
its power spectrum \cite{seljak96b,zaldarriagath} may lift to some extent some of these degeneracies.
It will also provide additional information on the reionization epoch and
on the presence of tensor perturbations, and may also help in the
identification and removal of polarized astrophysical foregrounds
\cite{kinney98,kamionkowski98,prunet99}.

A successful measurement of the CMB polarization stands as an
observational challenge; the expected polarization level is of the order of
$10\%$ of the level of temperature fluctuations ($\Delta T/T \simeq
10^{-5}$).  Efforts have thus gone into developing techniques to
reduce or eliminate spurious non-astronomical signals and instrumental
noise which could otherwise easily wipe out real polarization signals.
In a previous paper \cite{couchot98}, we have shown how to configure
the polarimeters in the focal plane in order to minimize the errors on
the measurement of the Stokes parameters. In this paper, we address the problem of
low frequency noise.

Low frequency noise in the data streams can arise due to a wide range
of physical processes connected to the detection of radiation.  $1/f$
noise in the electronics, gain instabilities, and temperature
fluctuations of instrument parts radiatively coupled to the detectors,
all produce low frequency drifts of the detector outputs.
\begin{figure}[ht]
  \begin{center}
    \epsfig{ file=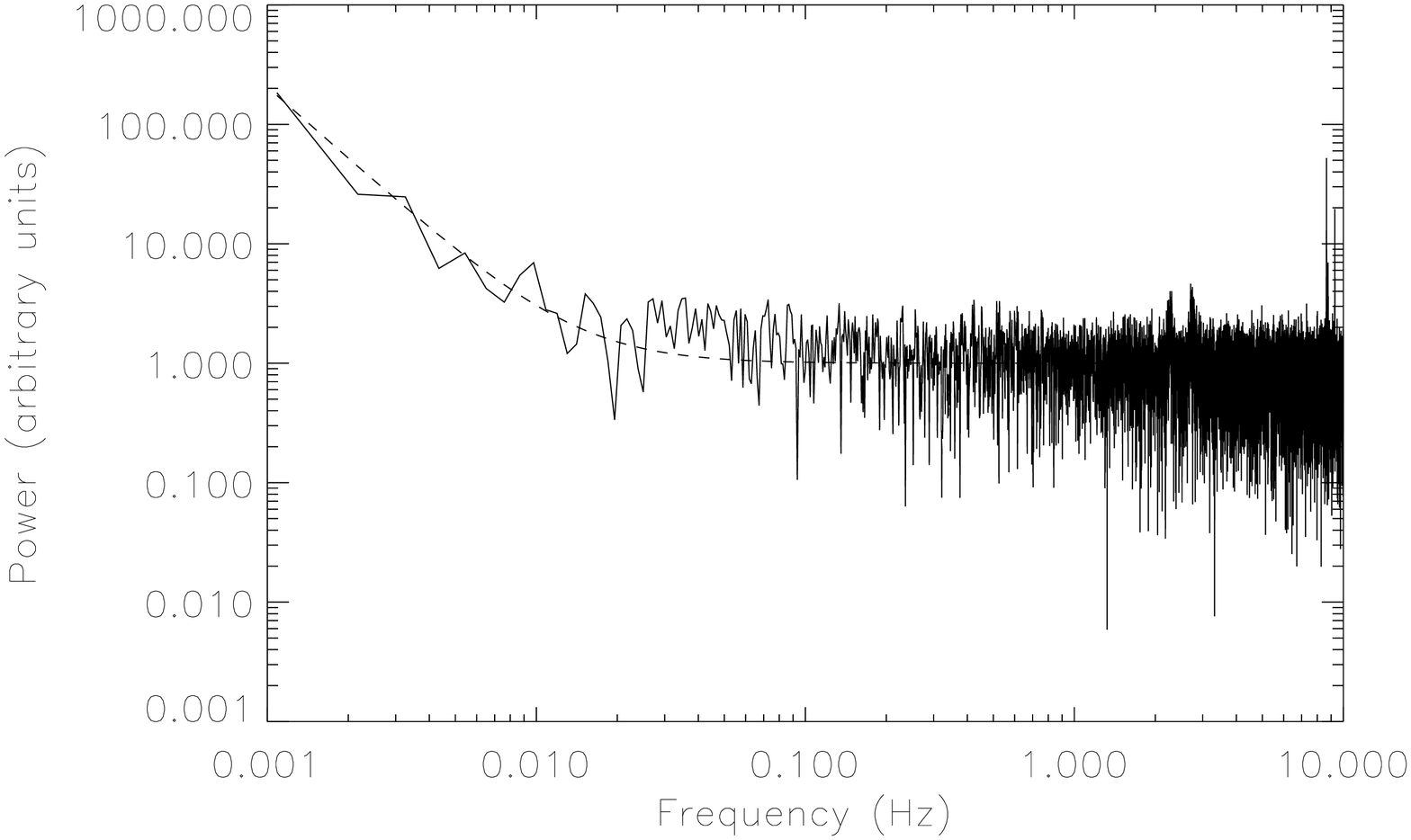,width=8.5cm}
  \end{center}
  \caption{The power spectrum of the K34 bolometer from Caltech, the same type of bolometer
    planned to be used on the {\sc Planck} mission. The measurement was performed at the SYMBOL test bench
    at I.A.S., Orsay (supplied by Michel Piat). The knee frequency of this spectrum is $\sim 0.014$ Hz 
    and the planned spin frequency for {\sc Planck} is $0.016$ Hz. 
    We can model the spectrum as the function
    ${S(f)=1+\left( \frac{1.43 \times 10^{-2}}{f} \right) ^2}$ (dashed line).
    \label{spectrebruit}}
\end{figure}
The spectrum of the total noise can be modeled as
a superposition of white noise and components behaving like
$1/f^{\alpha}$ where $\alpha \ge 1$, as shown in Fig.
\ref{spectrebruit}.

\begin{figure}[ht]
  \begin{center}
    \epsfig{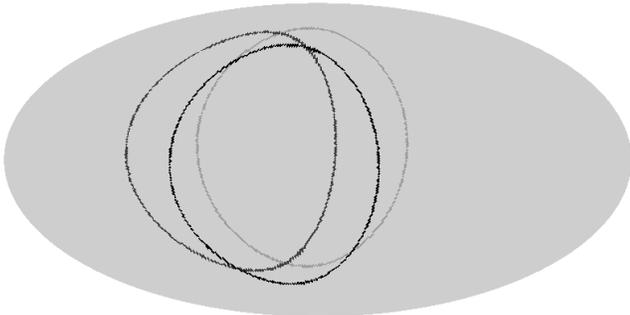}
  \end{center}
  \caption{The Mollweide projection of $3$ intersecting circles. For clarity, the scan angle 
    between the spin axis and the main beam axis is set to $60^\circ$ for this figure.
    \label{inter}}
\end{figure}

This noise generates stripes after reprojection on maps, whose exact form depends on the
scanning strategy. If not properly subtracted, the effect of such
stripes is to degrade considerably the sensitivity of an experiment.
The elimination of this ``striping'' may be achieved using
redundancies in the measurement, which are essentially of two types for
the case of {\sc Planck}:

\begin{itemize}
\item each individual detector's field of view scans the sky on large
  circles, each of which is covered consecutively many times ($\sim 60$) 
  at a rate of about ${f_\mathrm{spin}\sim 1}$~rpm.  This
  permits a filtering out of non scan-synchronous fluctuations in the
  circle constructed from averaging the consecutive scans.
\item a survey of the whole sky (or a part of it) involves many such
  circles that intersect each other (see Fig. \ref{inter}); the exact
  number of intersections depends on the scanning strategy but is of
  the order of $10^8$ for the {\sc Planck} mission: this will allow to
  constrain the noise at the intersection points.
\end{itemize}

\begin{figure}[ht]
  \begin{center}
    \epsfig{ file=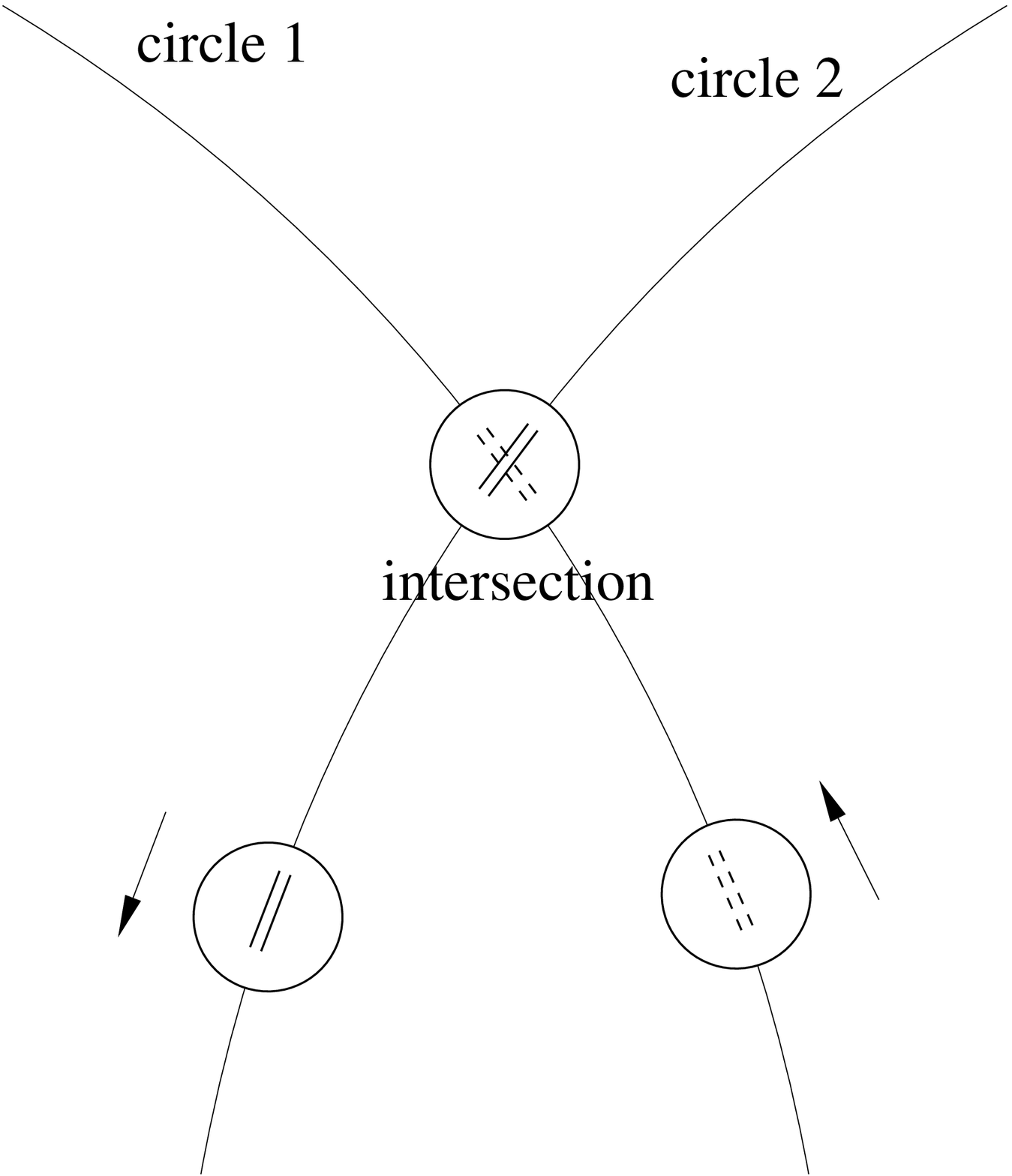, width=5cm}
  \end{center}
  \caption{
    The orientation of polarimeters at an intersection point. This
    point is seen by two different circles corresponding to two
    different orientations of the polarimeters in the focal plane.
    For clarity, we have just represented one polarimeter.
    \label{deuxpol}}
\end{figure}

One of us \cite{delabrouille98a} has proposed to remove low frequency drifts for
unpolarized data in the framework of the {\sc Planck} mission
by requiring that all measurements of a single
point, from all the circles intersecting that point, share a common
sky temperature signal. The problem is more complicated in the case of
polarized measurements since the orientation of a polarimeter with
respect to the sky depends on the scanning circle.  Thus, a given polarimeter
crossing a given point in the sky along two different circles will not
measure the same signal, as illustrated in Fig.  \ref{deuxpol}.

The rest of the paper is organized as follows: in Sect. \ref{noise},
we explain how we model the noise and how low frequency drifts
transform into offsets when considering the circles instead of
individual scans. In Sect. \ref{skypol}, we explain how polarization
is measured. The details of the algorithm for
removing low-frequency drifts are given in Sect. \ref{algo}. We
present the results of our simulations in Sect.  \ref{simul} and give
our conclusions in Sect. \ref{resul}.

\section{Averaging noise to offsets on circles}\label{noise}

As shown in Fig. \ref{spectrebruit}, the typical noise spectrum expected for
the {\sc Planck} High Frequency Instrument (HFI) features a drastic increase of noise
power at low frequencies ${f\leq 0.01 ~ \mathrm{Hz}}$. We model this noise spectrum
as:

\begin{equation}
\!\!\!\!\!\!\!\!\!  S(f) = \sigma^2 \times \left( 1 + \sum_i \left( \frac{f_i}{f} \right) ^{\alpha_i} \right).
\end{equation}

The knee frequency ${f_\mathrm{knee}}$ is defined as the frequency at which
the power spectrum due to low frequency contributions equals that of
the white noise. The noise behaves as pure white noise with variance
${\sigma^2}$ at high frequencies. The spectral index of each component
of the low-frequency noise, ${\alpha_i}$, is typically between 1 and 2,
depending on the physical process generating the noise.

\begin{figure}[ht]
  \begin{center}
    \epsfig{ file=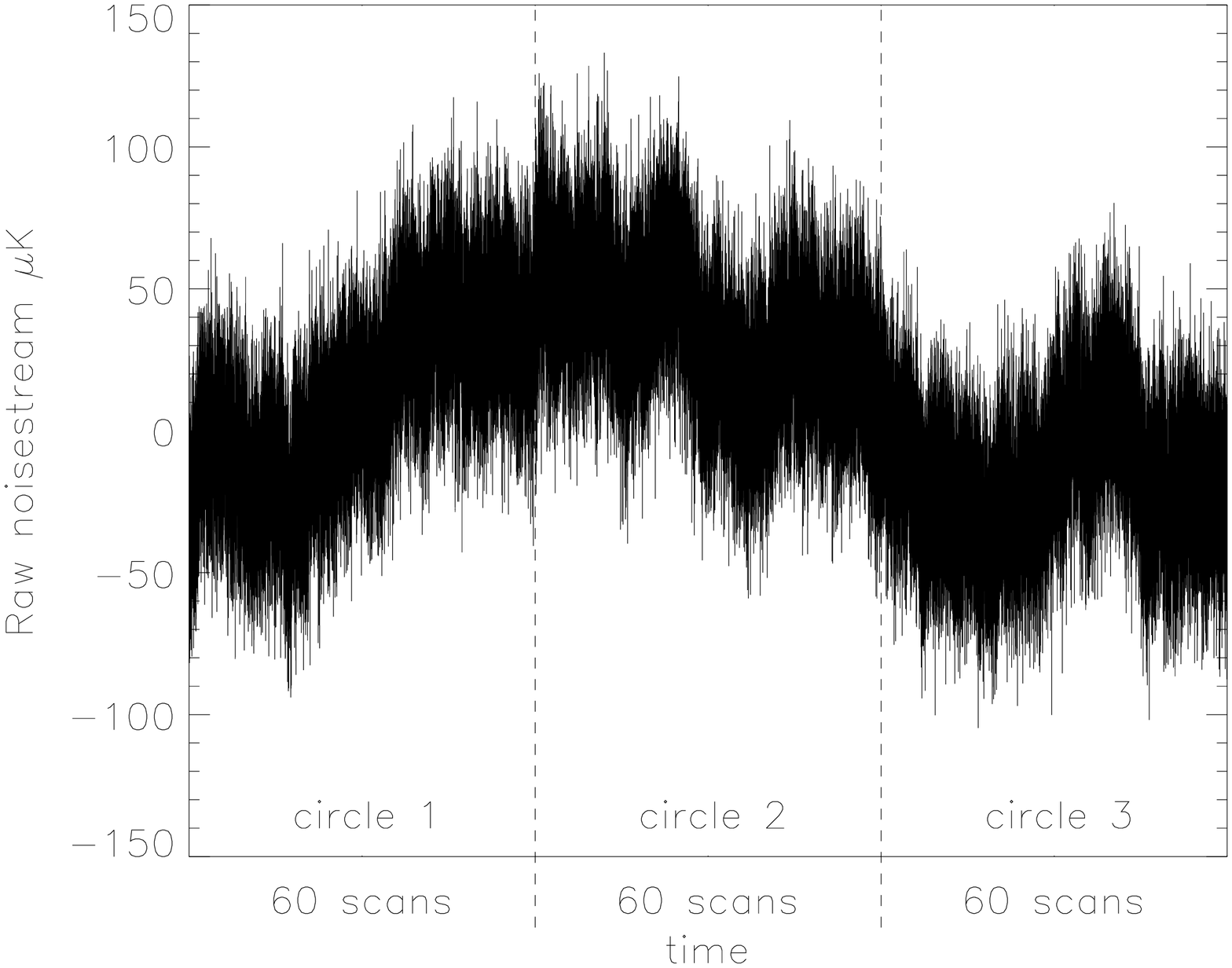, bbllx=25,bblly=20,bburx=675,bbury=530,width=8.5cm}
  \end{center}
  \caption{Typical $1/f^2$ low frequency noise stream.
    Here, ${f_{\rm knee}=f_{\rm spin}=0.016}$~Hz, ${\alpha=2}$ and
    ${\sigma=21 ~\mu\mathrm{K}}$ (see Eq. 1).
    This noise stream corresponds to 180 scans or 3 circles (60 scans per circle) or
    a duration of 3 hours.
    \label{stream1sf}}
\end{figure}

\begin{figure}[ht]
  \begin{center}
    \epsfig{ file=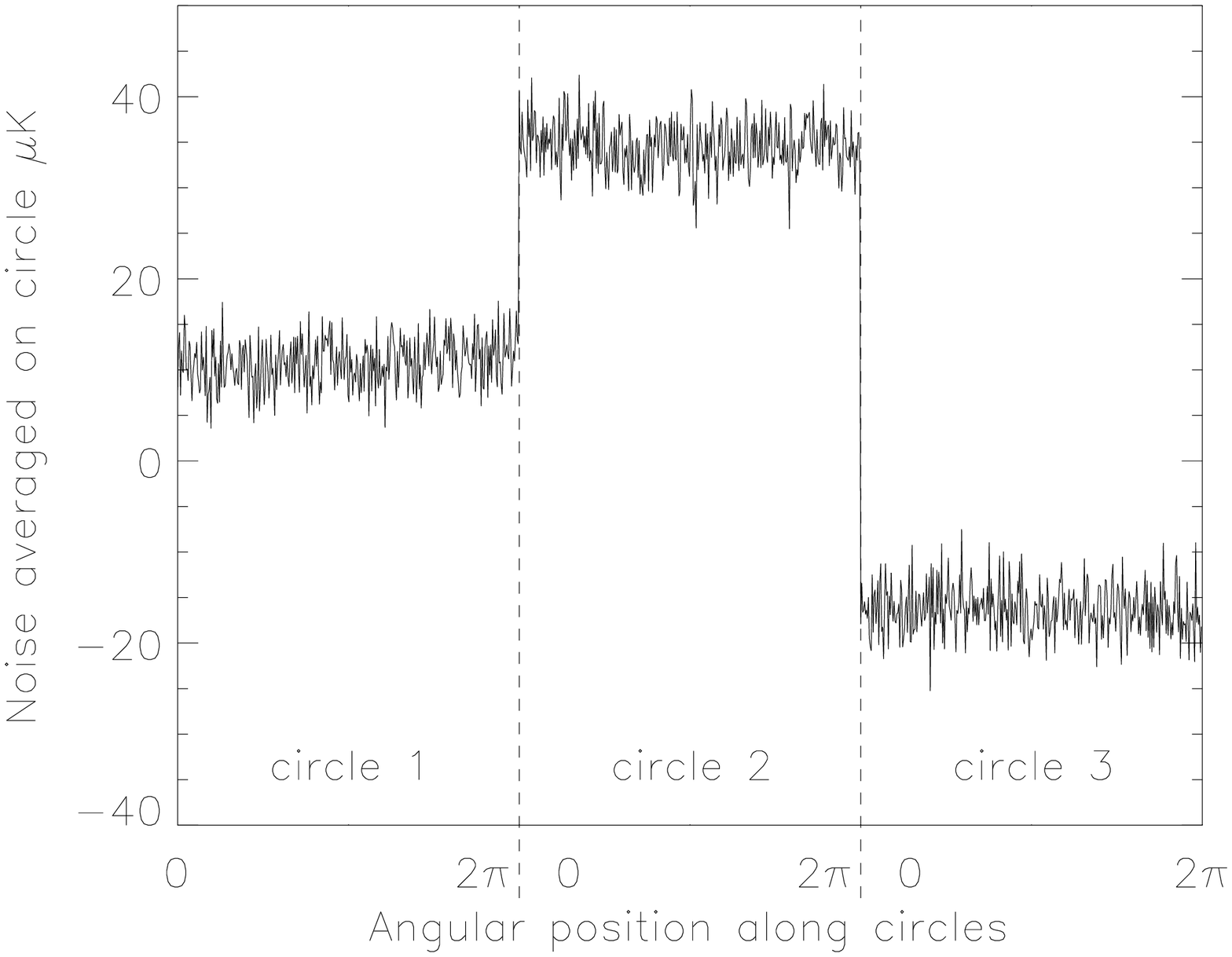, bbllx=55,bblly=50,bburx=655,bbury=515,width=8.5cm}
  \end{center}
  \caption{The residual noise on the 3 circles after averaging. To first approximation, 
    low frequency drifts are transformed into
    offsets, different for each circle and each polarimeter. Note the expanded scale on the $y$-axis as compared
    to that of Fig. \ref{stream1sf}.
    \label{average1sf}}
\end{figure}

The Fourier spectrum of the noise on the circle obtained by combining
$N$ consecutive scans depends on the exact method used. The simplest method, setting the circle
equal to the average of all its scans, efficiently filters out all
frequencies save the harmonics of the spinning frequency \cite{delabrouille97a}.  Since the
noise power mainly resides at low frequencies (see Fig.
\ref{spectrebruit}), the averaging transforms -- to first order -- low
frequency drifts into constant offsets different for each circle and
for each polarimeter. This is illustrated in the comparison between
Figs. \ref{stream1sf} and \ref{average1sf}. More sophisticated methods for recombining the data
streams into circles can be used, as $\chi^2$ minimization, Wiener filtering, or any map-making
method projecting about ${6\times 10^5}$ samples onto a circle of about ${5\times 10^3}$ points.
For simplicity, we will work in the following with the circles obtained by simple averaging of all its
consecutive scans.


We thus model the effect of low frequency drifts as a constant offset
for each polarimeter and each circle. This approximation is excellent
for ${f_\mathrm{knee}\le f_\mathrm{spin}}$. The remaining white noise of the $h$ polarimeters is
described by one constant ${h \times h}$ matrix.

\section{The measurement of sky polarization}\label{skypol}

\subsection{Observational method}

\begin{figure}[ht]
  \begin{center}
    \epsfig{ file=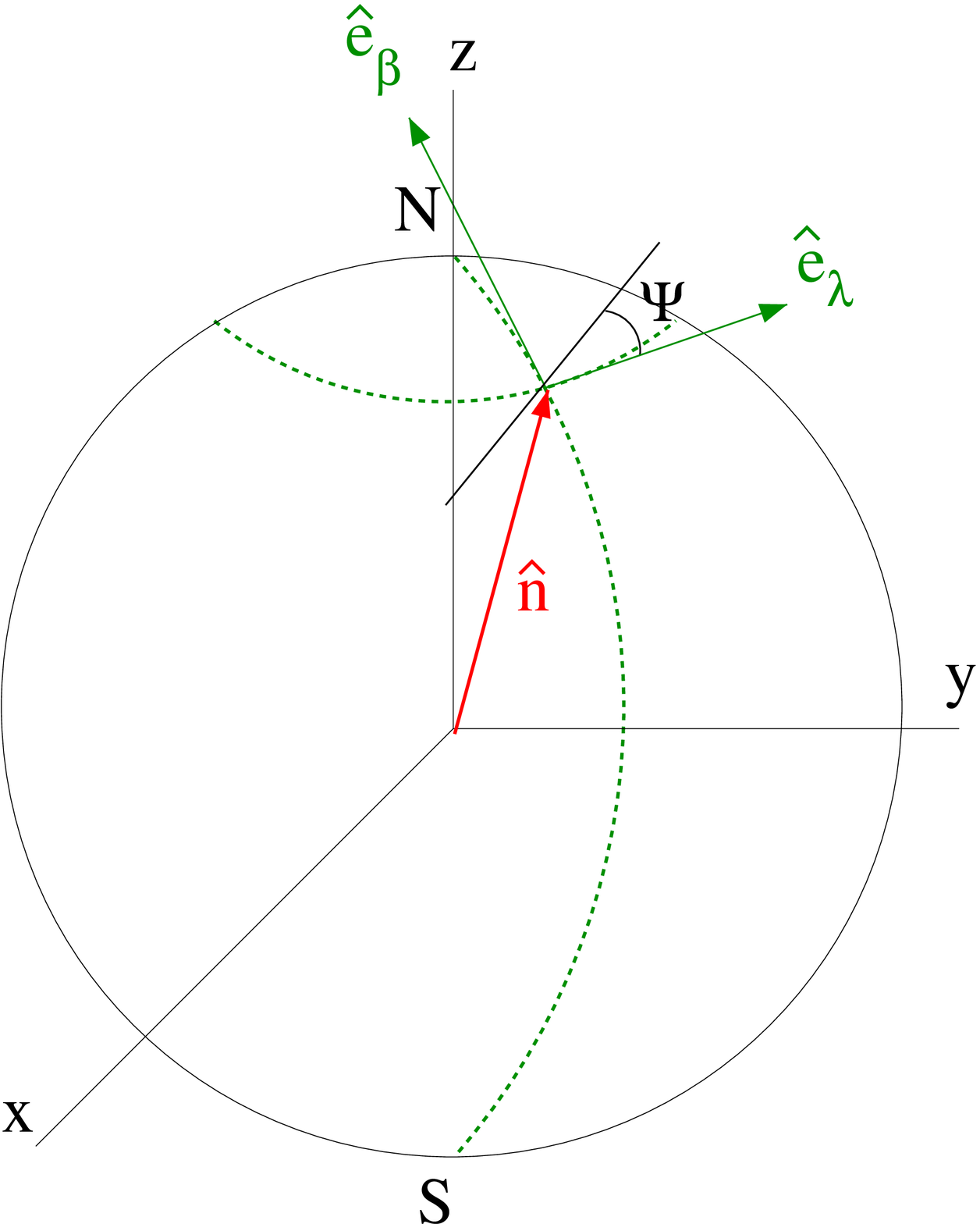,width=5cm}
  \end{center}
  \caption{The reference frame used to define the Stokes parameters and angular position $\Psi$
    of a polarimeter. $\Psi$ lies in the plane ${(\hat{e}_{\lambda},\hat{e}_{\beta})}$.
    \label{frame}}
\end{figure}

The measurement with one polarimeter of the linear polarization of a
wave coming from a direction $\boldsymbol{\hat{n}}$ on the sky, requires at least
three measurements with different polarimeter orientations. Since the
Stokes parameters  $Q$ and $U$ are not invariant under rotations, we define them
at each point $\boldsymbol{\hat{n}}$ with respect to a reference frame of tangential
vectors ${(\hat{e}_\lambda,\hat{e}_\beta)}$. The output signal given by a
polarimeter looking at point $\boldsymbol{\hat{n}}$ is:
\begin{equation}
\!\!\!\!\!\!\!\!\! M_{polar}=\frac{1}{2} ( I+Q \cos 2\Psi+U \sin 2\Psi ) 
\label{mesurepol}
\end{equation}
where $\Psi$ is the angle between the polarimeter and
$\hat{e}_\lambda$\footnote{We do not consider the $V$ Stokes parameter since
  no net circular polarization is expected through Thomson
  scattering.}.  In the following, we choose the longitude-latitude
reference frame as the fixed reference frame on the sky (see Fig. \ref{frame}).  

\subsection{Destriping method}

The destriping method consists in using redundancies at the
intersections between circle pairs to estimate, for each circle $i$
and each polarimeter $p$, the offsets $O_i^{p}$ on polarimeter
measurements. For each circle intersection, we require that all three
Stokes parameters  {\em in a fixed reference frame} in that direction of the sky, as
measured on each of the intersecting circles, be the same.  A $\chi^2$
minimization leads to a linear system whose solution gives the
offsets. By subtracting these offsets, we can recover the Stokes parameters  corrected
for low-frequency noise.

\subsection{Formalism}

We consider a mission involving $n$ circles.  The set of all circles
that intercept circle $i$ is denoted by $\mathcal{I}(i)$ and contains
$N_{\mathcal{I}(i)}$ circles.  For any pair of circles $i$ and $j$, we denote
the two points where these two circles intersect (if any) by
${\{i,j,\delta\}}$. In this notation $i$ is the circle currently
scanned, $j$ the intersecting circle in set $\mathcal{I}(i)$, and $\delta$
indexes the two intersections ($\delta = 1 (-1)$ indexes the first
(second) point encountered from the northernmost point on the circle)
so that the points ${\{j,i, -\delta\}}$ and ${\{i,j,\delta\}}$ on the sky
are identical.

The Stokes parameters  at point $\{i,j,\delta\}$, with respect to a fixed global
reference system, are denoted by a $3-$vector
$\boldsymbol{S}_{i,j,\delta}$, with
\begin{equation}
  \!\!\!\!\!\!\!\!\! \boldsymbol{S}_{i,j,\delta} = \boldsymbol{S}_{j,i,-\delta} = \left(
    \begin{array}{c}
      I\\
      Q\\
      U
      \end{array}
      \right)
      \left(
        \boldsymbol{\hat{n}} \equiv \{i,j,\delta \}
        \right)
      \label{redund}.
\end{equation}

At intersection $\{i,j,\delta\}$, the set of measurements by $h$
polarimeters travelling along the scanning circle $i$ is a $h-$vector
denoted by $\boldsymbol{M}_{i,j,\delta}$, and is related to the Stokes parameters  at this
point by (see Eq. \ref{mesurepol}):
\begin{equation}
  \!\!\!\!\!\!\!\!\! \boldsymbol{M}_{i,j,\delta} = \boldsymbol{\mathcal{A}}_{i,j,\delta} \boldsymbol{S}_{i,j,\delta} \label{stokesm}
\end{equation}
where $\boldsymbol{\mathcal{A}}_{i,j,\delta}$ is the $h \times 3$ matrix: 
\begin{equation}
  \!\!\!\!\!\!\!\!\!
  \boldsymbol{\mathcal{A}}_{i,j,\delta} =  \frac{1}{2}\left(\begin{array}{ccc}
      1&\cos 2\Psi_1(i,j,\delta)&\sin 2\Psi_1(i,j,\delta)\\
      \vdots & \vdots & \vdots\\
      1&\cos 2\Psi_p(i,j,\delta)&\sin 2\Psi_p(i,j,\delta)\\
      \vdots & \vdots & \vdots\\
      1&\cos 2\Psi_h(i,j,\delta)&\sin 2\Psi_h(i,j,\delta)\\
    \end{array}\right).
  \nonumber
\end{equation}

$\Psi_p(i,j,\delta) \in [0,\pi]$ is the angle between the orientation of
polarimeter $p$ and the reference axis in the fixed global reference
frame (see Fig. \ref{frame}). The matrix $\boldsymbol{\mathcal{A}}_{i,j,\delta}$ can be
factorised as
\begin{equation}
  \!\!\!\!\!\!\!\!\! \boldsymbol{\mathcal{A}}_{i,j,\delta} = \boldsymbol{\mathcal{A}}  \boldsymbol{R}_{i,j,\delta}. \label{decomp}
\end{equation}
The constant $h \times 3$ matrix $\boldsymbol{\mathcal{A}}$ characterizes the geometrical
setup of the $h$ polarimeters in the focal reference frame:
\begin{equation}
  \!\!\!\!\!\!\!\!\!\boldsymbol{\mathcal{A}} =\frac{1}{2} \left(\begin{array}{ccc}
      1&1&0\\
      \vdots & \vdots & \vdots\\
      1&\cos 2 \Delta_p&\sin 2 \Delta_p\\
      \vdots & \vdots & \vdots\\
      1&\cos 2 \Delta_h&\sin 2 \Delta_h\\
    \end{array}\right)\label{matabsA}
\end{equation}
where $\Delta_p$ is the angle between the orientations of polarimeters
$p$ and $1$, so we have $\Psi_p=\Psi_1+\Delta_p$ and $\Delta_1=0$.
The rotation matrix $\boldsymbol{R}_{i,j,\delta}$ brings the focal plane to its
position at intersection $\{i,j,\delta\}$ when scanning along circle
$i$:
\begin{equation}
  \!\!\!\!\!\!\!\!\!\boldsymbol{R}_{i,j,\delta}=
  \left(
    \begin{array}{ccc}
      1&0&0\\
      0&\cos 2\Psi_1(i,j,\delta)&\sin 2\Psi_1(i,j,\delta)\\
      0&-\sin 2\Psi_1(i,j,\delta)&\cos 2\Psi_1(i,j,\delta) 
    \end{array}\right). \label{arot}
\end{equation} 

\section{The algorithm}\label{algo}
\subsection{The general case}

To extract the offsets from the measurements, we use a $\chi^2$
minimization.  This $\chi^2$ relates the measurements $\boldsymbol{M}_{i,j,\delta}$
to the offsets $\boldsymbol{O}_i$ and the Stokes parameters  $\boldsymbol{S}_{i,j,\delta}$, using the
redundancy condition \eqref{redund}. In order to take into account the
two contributions of the noise (see Sect. \ref{noise}) and of the
Stokes parameters  (see Eq. \ref{stokesm}), we model the measurement as:
\begin{equation}\label{eq:offsetsmodel}
  \!\!\!\!\!\!\!\!\! \boldsymbol{M}_{i,j,\delta}=
  \boldsymbol{\mathcal{A}} \, \boldsymbol{R}_{i,j,\delta} \, \boldsymbol{S}_{i,j,\delta} \, + \boldsymbol{O}_i \, + \mathrm{white~noise}.
\end{equation}
so that we write
\begin{multline}
  \!\!\!\!\!\!\!\!\! \chi^2 = \sum_{i,j\in {\mathcal{I}(i)},\delta=\pm 1} \,\left(\boldsymbol{M}_{i,j,\delta} - \boldsymbol{O}_i -
    \boldsymbol{\mathcal{A}}\,\boldsymbol{R}_{i,j,\delta}\, \boldsymbol{S}_{i,j,\delta}\right)^T \times\\
  {\boldsymbol{N}_i}^{-1}\left(\boldsymbol{M}_{i,j,\delta} - \boldsymbol{O}_i -
    \boldsymbol{\mathcal{A}}\,\boldsymbol{R}_{i,j,\delta}\, \boldsymbol{S}_{i,j,\delta}\right).
\label{eq:chi2}
\end{multline}
where ${\boldsymbol{N}_i}$ is the $h \times h$ matrix of noise correlation
between the $h$ polarimeters. 

Minimization with respect to $\boldsymbol{O}_i$ and $\boldsymbol{S}_{i,j,\delta}$ yields the
following equations: 
\begin{gather}
  \!\!\!\!\!\!\!\!\!  {\boldsymbol{N}_i}^{-1} \sum_{j\in \mathcal{I}(i),\delta=\pm 1} \left(\boldsymbol{M}_{i,j,\delta} - \boldsymbol{O}_i -
    \boldsymbol{\mathcal{A}}\,\boldsymbol{R}_{i,j,\delta}\, \boldsymbol{S}_{i,j,\delta}\right) = 0, \label{eqgd}
  \!\!\!\!\!\!\!\!\!  \intertext{and}
  \!\!\!\!\!\!\!\!\!  \boldsymbol{R}_{i,j,\delta}^{-1}\,\boldsymbol{\mathcal{A}}^T\, {\boldsymbol{N}_i}^{-1} \left(\boldsymbol{M}_{i,j,\delta} - \boldsymbol{O}_i -
    \boldsymbol{\mathcal{A}}\,\boldsymbol{R}_{i,j,\delta}\, \boldsymbol{S}_{i,j,\delta}\right) + \notag\\
  \!\!\!\!\!\!\!\!\! \boldsymbol{R}_{j,i,-\delta}^{-1} \boldsymbol{\mathcal{A}}^T {\boldsymbol{N}_j}^{-1} \left(\boldsymbol{M}_{j,i,-\delta} - \boldsymbol{O}_j -
      \boldsymbol{\mathcal{A}}\,\boldsymbol{R}_{j,i,-\delta}\, \boldsymbol{S}_{i,j,\delta}\right) = 0.
  \label{eqgs}
\end{gather}

We can work with a reduced set of transformed measurements and
offsets which can be viewed as the Stokes parameters  in the focal reference
frame and the associated offsets which are the $3$ dimensional vectors:
\begin{gather}
  \!\!\!\!\!\!\!\!\! \mathscr{S}_{i,j,\delta} = {\boldsymbol{X}_i}^{-1}\,\boldsymbol{\mathcal{A}}^T\, {\boldsymbol{N}_i}^{-1} \boldsymbol{M}_{i,j,\delta} \notag \mathrm{~~and~}\\
  \!\!\!\!\!\!\!\!\! \boldsymbol{\Delta}_i = {\boldsymbol{X}_i}^{-1}\,\boldsymbol{\mathcal{A}}^T\, {\boldsymbol{N}_i}^{-1} \boldsymbol{O}_i, \label{defg}\\ 
  \!\!\!\!\!\!\!\!\! \mbox{where } \boldsymbol{X}_i = \boldsymbol{\mathcal{A}}^T\, {\boldsymbol{N}_i}^{-1}\, \boldsymbol{\mathcal{A}}. \notag 
\end{gather}
Eqs. \eqref{eqgd} and \eqref{eqgs} then simplify to:
\begin{gather}
  \!\!\!\!\!\!\!\!\! \sum_{j\in \mathcal{I}(i),\,\delta=\pm 1} \left(\mathscr{S}_{i,j,\delta} - \boldsymbol{\Delta}_i - \boldsymbol{R}_{i,j,\delta}\,
    \boldsymbol{S}_{i,j,\delta}\right) = 0, \label{eqgd2}
  \!\!\!\!\!\!\!\!\! \intertext{and}
  \!\!\!\!\!\!\!\!\! \boldsymbol{R}_{i,j,\delta}^{-1}\,\boldsymbol{X}_i\,\left(\mathscr{S}_{i,j,\delta} - \boldsymbol{\Delta}_i -
    \boldsymbol{R}_{i,j,\delta}\, \boldsymbol{S}_{i,j,\delta}\right) + \notag\\
  \!\!\!\!\!\!\!\!\! \boldsymbol{R}_{j,i,-\delta}^{-1}\,\boldsymbol{X}_j\ \left(\mathscr{S}_{j,i,-\delta}
    - \boldsymbol{\Delta}_j - \boldsymbol{R}_{j,i,-\delta}\, \boldsymbol{S}_{i,j,\delta}\right) = 0. 
  \label{eqgs2}
\end{gather}
$\boldsymbol{R}_{i,j,\delta}\,\boldsymbol{S}_{i,j,\delta}$ in Eq. \eqref{eqgs2} can be solved for
and the result inserted in Eq. \eqref{eqgd2}. After a few
algebraic manipulations, one gets the following linear system for the
offsets $\boldsymbol{\Delta}_i$ as functions of the data $\mathscr{S}_{i,j,\delta}$:
\begin{multline}
  \!\!\!\!\!\!\!\!\!  \sum_{j\in \mathcal{I}(i),\delta=\pm 1}\left[\bbbone + \widetilde{\boldsymbol{R}}(i,j,\delta)\,
    {\boldsymbol{X}_j}^{-1}\,\widetilde{\boldsymbol{R}}(i,j,\delta)^{-1}\,\boldsymbol{X}_i\right]^{-1} \times \\
  \!\!\!\!\!\!\!\!\!\!\!\!\!\!\!\!\!\!  \left[\boldsymbol{\Delta}_i - \widetilde{\boldsymbol{R}}(i,j,\delta)\,\boldsymbol{\Delta}_j - 
    \mathscr{S}_{i,j,\delta} + \widetilde{\boldsymbol{R}}(i,j,\delta)\,\mathscr{S}_{j,i,-\delta}\right]  = 0,
  \label{eqgd3}
\end{multline}
where the rotation
\begin{equation}
  \!\!\!\!\!\!\!\!\!\widetilde{\boldsymbol{R}}(i,j,\delta)=\boldsymbol{R}_{i,j,\delta} \,{\boldsymbol{R}^{-1}_{j,i,-\delta}} \label{frot}
\end{equation}
brings the focal reference frame from its position along scan $j$ at
intersection $\{i,j,\delta\}$ to its position along scan $i$ at the same
intersection (remember that $\{i,j,\delta\} = \{j,i,-\delta\}$). 
Note that
$\widetilde{\boldsymbol{R}}(i,j,\delta)=\widetilde{\boldsymbol{R}}(j,i,-\delta)^{-1}$. 
In this linear system, we need to know the measurements of the
polarimeters
at the points $\{i,j,\delta \}$ and $\{j,i,-\delta\}$. These two points on
circles $i$ and $j$ respectively will unlikely correspond to a
sample along these circles. So we have linearly interpolated the
value of the intersection points from the values measured at sampled points. 
For a fixed circle $i$, this is a $3 \times N_{\mathcal{I}(i)}$ linear
system. In Eq. \eqref{eqgd3}, $i$ runs from $1$ to $n$, therefore the
total matrix to be inverted has dimension $3n \times 3n$.  However,
because the rotation matrices are in fact two dimensional (see Eq.
\ref{arot}), the intensity components $\Delta_i^I$ of the offsets
only enter Eq. \eqref{eqgd3} through their differences $\Delta_i^I -
\Delta_j^I$ so that the linear system is not invertible: the rank of
the system is $3n-1$.  In order to compute the offsets, we can fix the
intensity offset on one particular scanning circle or add the
additionnal constraint that the length of the solution vector is
minimized.

Once the offsets $\boldsymbol{\Delta}_i$ are known, 
the Stokes parameters  in the global reference frame
$\left( \hat{e}_{\lambda},\hat{e}_{\beta} \right)$ at a generic
sampling $k$ of the circle
$i$, labeled by $\{i,k\}$  are estimated as
\begin{equation}
  \!\!\!\!\!\!\!\!\!\boldsymbol{S}_{i,k} =  \boldsymbol{R}_{i,k}^{-1} \left(\mathscr{S}_{i,k} - \boldsymbol{\Delta}_i\right),
  \label{substract}
\end{equation}
where $\boldsymbol{R}_{i,k}$ is the rotation matrix which transforms the focal frame Stokes parameters  into those of
the global reference frame.

The quantities $\mathscr{S}_{i,k}$ are the Stokes parameters  measured in the focal
frame of reference at this point and are simply given in terms of the
measurements $\boldsymbol{M}_{i,k}$ (see Eq. \ref{defg}) by:
\begin{equation}
  \!\!\!\!\!\!\!\!\! \mathscr{S}_{i,k} = {\boldsymbol{X}_i}^{-1}\,\boldsymbol{\mathcal{A}}^T\, {\boldsymbol{N}_i}^{-1} \boldsymbol{M}_{i,k}.
\end{equation}

The matrix
\begin{equation}
  \!\!\!\!\!\!\!\!\! \boldsymbol{N}^{\rm{Stokes}}_i=\left( \boldsymbol{\mathcal{A}}^T \, \boldsymbol{N}_i^{-1} \, \boldsymbol{\mathcal{A}} \right) ^{-1}
\end{equation}
is the variance matrix of the Stokes parameters  on circle $i$.  Note that this
algorithm is totally independent of the pixelization chosen which only
enters when reprojecting the Stokes parameters  on the sphere.

\subsection{Uncorrelated polarimeters, with identical noise}
When the polarimeters are uncorrelated with identical noise, the
variance matrix reduces to $\boldsymbol{N}_i = \bbbone/{\sigma_i}^2$ and the matrices
$\boldsymbol{X}_i$ can all be written as
\[
\!\!\!\!\!\!\!\!\! \boldsymbol{X}_i = \frac{1}{{\sigma_i}^2}\,\boldsymbol{X} \mbox{ with } \boldsymbol{X} = \boldsymbol{\mathcal{A}}^T
\boldsymbol{\mathcal{A}}
\]

\subsection*{Case of ``Optimized Configurations''}
We have shown \cite{couchot98} that the polarimeters can be 
arranged in ``Optimized Configurations'', where the $h$ polarimeters are
separated by angles of $\pi/h$. If the noise level of each of the $h$
polarimeters is the same and if there are no correlation between detector noise, 
then the errors of the
Stokes parameters  are also decorrelated and the matrix $\boldsymbol{X}$ has the simple form:
\begin{equation}
\!\!\!\!\!\!\!\!\!
\boldsymbol{X} = \frac{n}{8}\left(
  \begin{array}{ccc}
    2&0&0\\
    0&1&0\\
    0&0&1
  \end{array}\right).
\end{equation}
Because this matrix commutes with all rotation matrices
$\widetilde{\boldsymbol{R}}(i,j,\delta)$, 
Eq. \eqref{eqgd3} simplifies further to
\begin{multline}
  \!\!\!\!\!\!
  \frac{N_{\mathcal{I}(i)}}{\sigma_{\mathcal{I}(i)}^2}\boldsymbol{\Delta}_i
  - \sum_{{m\in
      {\mathcal{I}(i)}}}\frac{1}{\sigma_i^2+\sigma_j^2}\left(\widetilde{\boldsymbol{R}}(i,j,1)+\widetilde{\boldsymbol{R}}(i,j,-1)
  \right)\boldsymbol{\Delta}_j\\
  \!\!\!\!\!\!\!\!\! = \sum_{j\in {\mathcal{I}(i)},\delta=\pm 1}
  \frac{1}{\sigma_i^2+\sigma_j^2} \left(\mathscr{S}_{i,j,\delta} -
    \widetilde{\boldsymbol{R}}(i,j,\delta) \,
    \mathscr{S}_{j,i,-\delta}\right),
  \label{eqdelta1}
\end{multline}
where the sum over $\delta$ is explicit on the left side of the
equation and we have defined an average error
$\sigma_{\mathcal{I}(i)}$ along circle $i$ by
\[
\!\!\!\!\!\!\!\!\!\frac{N_{\mathcal{I}(i)}}{\sigma_{\mathcal{I}(i)}^2} = \sum_{j\in{\mathcal{I}(i)}} \frac{2}{\sigma_i^2+\sigma_j^2},
\]
and where the rotation matrix $\widetilde{\boldsymbol{R}}(i,j,\delta)$ is defined by Eq. \eqref{frot}.
Rotations $\widetilde{\boldsymbol{R}}(i,j,\delta)$ and
$\widetilde{\boldsymbol{R}}(i,j,-\delta)=\widetilde{\boldsymbol{R}}(j,i,\delta)^{-1}$ correspond to the two intersections
between circles $i$ and $j$. Eq. \eqref{eqdelta1} can be simplified 
further. We can separate the $\boldsymbol{\Delta}_i$ and the $\mathscr{S}_{i,j,\delta}$ into
scalar components related to the intensity: $\Delta^I_i,\
\mathscr{S}^I_{i,j,\delta}$ and  2-vectors components related to the
polarization: $\boldsymbol{\Delta}^P_i,\ \mathscr{S}^P_{i,j,\delta}$. We obtain then two separate equations, one for the 
intensity offsets $\Delta^I_i$, which is exactly the same as in the
unpolarized case (see \ref{appendixa}):
\begin{multline}
\!\!\!\!\!\! \sum_{j\in {\mathcal{I}(i)}}\frac{2}{\sigma_i^2+\sigma_j^2}\,\left( \Delta^I_i -
  \Delta^I_j\right) = \\
\!\!\!\!\!\!\!\!\! \qquad \qquad\sum_{j\in {\mathcal{I}(i)},\delta=\pm 1}
\frac{1}{\sigma_i^2+\sigma_j^2}\left(\mathscr{S}^I_{i,j,\delta} 
  - \mathscr{S}^I_{j,i,-\delta}\right),
\label{eqdeltaI}
\end{multline}
and one for the polarization offsets $\boldsymbol{\Delta}^P_i$:
\begin{multline}
\!\!\!\!\!\! \frac{N_{\mathcal{I}(i)}}{\sigma_{\mathcal{I}(i)}^2}\,\boldsymbol{\Delta}^P_i - 
\sum_{j\in {\mathcal{I}(i)}}\frac{2}{\sigma_i^2+\sigma_j^2}\, \cos
\left(
  2 
    \Psi_{ij}
\right)
\,\boldsymbol{\Delta}^P_j = \\
\!\!\!\!\!\!\!\!\! \sum_{j\in {\mathcal{I}(i)},\delta=\pm 1}\frac{1}{\sigma_i^2+\sigma_j^2} 
\left(\mathscr{S}^P_{i,j,\delta} - \boldsymbol{\mathcal{R}}(i,j,\delta)\,\mathscr{S}^P_{j,i,-\delta}\right),
\label{eqdeltaP}
\end{multline}
where $\Psi_{ij}=\Psi_1(i,j,\delta)-\Psi_1(j,i,-\delta)$ is the angle of the rotation that
brings the focal reference frame along scan $j$ on the focal reference
frame along scan $i$ at intersection $\{i,j,\delta\}$. The two
dimensional matrix $\boldsymbol{\mathcal{R}}(i,j,\delta)$ is the rotation sub-matrix
contained in $\widetilde{\boldsymbol{R}}(i,j,\delta)$ (see Eq. \ref{frot}).
Note that all mixing between polarization components have disappeared
from the left side of Eq. \eqref{eqdeltaP}). Therefore we are
left with two different $n \times n$ matrices to invert 
in order to solve for the offsets $\boldsymbol{\Delta}_i$, instead of one $3n \times
3n$ matrix.

As in Eq. \eqref{eqgd3}, the linear system in Eq. \eqref{eqdeltaI}
involves differences between the offsets $\Delta_i^I$, the matrix is
not invertible, and we find the solution in the same way as in the
general case.  On the other hand, for the polarized offsets
$\Delta_i^P$, the underlying matrix of Eq. \eqref{eqdeltaP} is regular
as expected.

\section{Simulations and test of the algorithm}\label{simul}

\subsection{Methods}
We now discuss how we test the method using numerical simulations.
For each simulated mission, we produce several maps.  The first,
which we use as the standard
``reference'' of comparison, is a projected map of a mission
with only white noise, ${f_\mathrm{knee}=0}$.
The remaining maps include $1/f$ plus white noise streams
with ${f_\mathrm{knee}=\eta f_\mathrm{spin}}$, ${\eta \in \{1,2,5,10\}}$.
The first of these is an ``untreated'' image which is projected with
no attempt to remove striping affects.  In the second
``zero-averaged'' map, we attempt a crude destriping by subtracting its average to each circle.
The final ``destriped'' map is constructed using the algorithm
in this paper.
We subtract the input maps ($I$, $Q$ and $U$) from the final maps in order to get maps of noise residuals.
Note that in case of a zero signal sky, setting the average of each circle to zero is better than destriping by nature 
because the offsets are only due to the noise. With a real sky,
both signal and noise contribute to the average so that zeroing circle not only removes the noise but also the signal.
Giard~et~al~\cite*{giard99} have attempted to refine their method by fitting templates of the dipole and of the galaxy
before subtracting a baseline from each circle. They concluded that an additional destriping (they used the algorithm of
Delabrouille~\cite*{delabrouille98a}) is needed.

\subsection{Simulated missions}
In order to test its efficiency, the destriping algorithm has been
applied to raw data streams generated from simulated observations
using various circular scanning strategies representative of a satellite mission as
{\sc Planck}, different ``Optimized Configurations'', and various noise
parameters.  The resulting maps were then compared with input and untreated
maps to test the quality of the destriping.

The input temperature ($I$) maps are the sum of galaxy, dipole,
and a randomly generated standard CDM anisotropy map 
(we used HEALPIX\footnote{\tt http://www.tac.dk/\~{}healpix/}
and CMBfast\footnote{\tt http://www.sns.ias.edu/\~{}matiasz\\
  /CMBFAST/cmbfast.html}).
Similarly, the polarization maps $Q$ and $U$ are the sum of the galaxy and
CMB polarizations.
The CMB polarization maps are randomly generated assuming a
standard CDM scenario. For the galactic polarization maps, we constructed a
random, continuous and correlated vector field defined on the 2-sphere with a correlation length of $5^\circ$
and a maximum polarization rate of ${20\%}$ ($100\%$ gave similar results).
Given the temperature map of the sky (not including CMB contribution), we can thus construct two polarization
maps for $Q$ and $U$.

\subsection{Results}
We first consider the case of destriping pure white noise and check
that the destriping algorithm does not introduce spurious structure.
Once this is verified, we apply the destriping algorithm to low
frequency noise. We find that the quality of the destriping is significantly dependent
on $\eta$ only. 
To demonstrate visually the quality of the destriping, we produce projected
sky maps with the input galaxy, dipole and CMB signal subtracted.

For temperature maps, we can compare Figs. 7,8, and 9. The eye is not able to see any differences
between the white noise map and the noise residual on the destriped map. 
We will see in the following how to quantify the presence of structures.
For the ``zero-averaged circles'' map, the level of the structure is very high and make it impossible
to compute the power spectrum of the CMB (see Fig. 13).

For the $Q$ Stokes parameter, Fig. 11 shows the destriped map for $\eta=1$.
Fig. 12 shows a map where the offsets are calculated as the average of
each circle. The maps for $U$ are very similar.
As for the $I$ maps, the destriped map is very similar to the white noise map. There exist some residual
structure on the ``zero-averaged circles'' map.
To assess quantitatively the efficiency of the destriping algorithm,
we have first studied the power spectra $C_\ell^T$, $C_\ell^E$, $C_\ell^B$, $C_\ell^{TE}$ and $C_\ell^{TB}$
calculated from the $I$, $Q$ and $U$ maps
\cite{zaldarriaga97,kamionkowski97}.
\begin{figure}[ht]
  \caption{The Mollweide projection of the residuals of the 
    ${I-\mathrm{Stokes}}$ parameter for a white noise mission. 
    The scale is in Kelvins. 
    The parameters of the simulation leading to this map are 
    described in \ref{appendixb}.
    \label{siwun}}
\end{figure}
\begin{figure}[ht]
  \caption{The Mollweide projection of the residuals of the 
    ${I-\mathrm{Stokes}}$ parameter after destriping of $1/f$ noise plus white noise with $\eta=1$
    \label{sidun}}
\end{figure}
\begin{figure}[ht]
  \caption{The Mollweide projection of the residuals of the 
    ${I-\mathrm{Stokes}}$ parameter after zeroing the average of the circles, 
    for $1/f$ noise plus white noise with $\eta=1$.
    \label{simun}}
\end{figure}
\begin{figure}[ht]
  \caption{The Mollweide projection of the residuals of the 
    ${Q-\mathrm{Stokes}}$ parameter for a white noise mission.
    \label{sqwun}}
\end{figure}
\begin{figure}[ht]
  \caption{The Mollweide projection of the residuals of the ${Q-\mathrm{Stokes}}$
    parameter after destriping of $1/f$ noise plus white noise with $\eta=1$
    \label{sqdun}}
\end{figure}
\begin{figure}[ht]
  \caption{The Mollweide projection of the residuals of the ${Q-\mathrm{Stokes}}$
    parameter after zeroing the average of the circles, for $1/f$ noise plus white noise with $\eta=1$.
    Although the remaining structures seem small, they are responsible for the excess of power in 
    ${C_\ell^E}$, see Fig. \ref{speun}.
    \label{sqmun}}
\end{figure}

The reference sensitivity of our simulated mission is evaluated by
computing the average spectra of 1000 maps of reprojected mission white noise.
This reference sensitivity falls, within sample variance, between the two dotted lines
represented in Figs. 13, 14, 15 and 16 and below the
dotted line in Figs. 17 and 18.
Figs. 13 and 14 show the spectra ${C_\ell^T}$ 
corresponding to ${f_\mathrm{knee}/f_\mathrm{spin}=1\mathrm{~and~}5}$ respectively,
for the $T$ field.
Similarly, Figs. 15 and 16 are the spectra ${C_\ell^E}$.
The $B$ field is not represented because it is very similar to $E$.
Figs. 17 and 18 represent the correlation between $E$ and $T$:~$C_\ell^{ET}$.
For ${f_\mathrm{knee}/f_\mathrm{spin}=1}$, we see that we are able to remove very efficiently low frequency
drifts in the noise stream: the destriped spectra obtained are
compatible with the white spectrum (within sample variance). Similar quality destriping is
achieved for any superposition of $1/f$, $1/f^2$ and white noise,
provided that the knee frequency is lower than or equal to the spin
frequency.
In the case of ${f_\mathrm{knee}/f_\mathrm{spin}=5}$, 
the method as implemented here leaves some
striping noise on the maps at low values of $\ell$.  Modeling the
noise as an offset is no longer adequate and a better model of the
averaged low-frequency noise is required (superposition of sine and
cosine functions for instance), or a more sophisticated method for constructing
one circle from 60 scans.  We again note that the value of the
ratio ${f_\mathrm{knee}/f_\mathrm{spin}}$ for both {\sc Planck}  HFI and LFI is likely to be very close
to unity (in Fig.  \ref{spectrebruit}, ${f^\mathrm{measured}_\mathrm{knee}\sim 0.014}$~Hz and 
${f_\mathrm{spin}=0.016}$~Hz).

To quantify the presence of stripes in the maps of residuals, we can compute the value of the ``striping'' estimator
${\mathrm{rms}(a^{T,E,B}_{\ell \ell})/\mathrm{rms}(a^{T,E,B}_{\ell 0})}$, because stripes tend to appear as structure
grossly parallel to the iso-longitude circles. In the case of pure white noise with a uniform
sky coverage, this value is 1. Here, because of the scanning strategy, the sky coverage
is not uniform and the value of this estimator is greater than 1, showing that it is not
specific of the striping. In order to get rid of the effect of non-uniform sky coverage, we express the estimator
${\mathrm{rms}(a^{T,E,B}_{\ell \ell})/\mathrm{rms}(a^{T,E,B}_{\ell 0})}$
in units of ${\mathrm{rms}(a^{T,E,B}_{\ell \ell})/\mathrm{rms}(a^{T,E,B}_{\ell 0})}$ for the white noise. This new estimator
is specific to the striping. The results in Table \ref{table1} show the improvement achieved by the destriping
algorithm although the result is still not perfect.
\begin{table}\caption{Values of ${\mathrm{rms}(a^{T,E,B}_{\ell \ell})/\mathrm{rms}(a^{T,E,B}_{\ell 0})}$ as
    a function of ${\eta={f_\mathrm{knee}/f_\mathrm{spin}}}$ in units of 
    ${\mathrm{rms}(a^{T,E,B}_{\ell \ell}(\mathrm{WN}))/\mathrm{rms}(a^{T,E,B}_{\ell 0}(\mathrm{WN}))}$
    We have checked that the systematic difference between the zero-averaged $E$ and $B$ fields is randomly in favor
    of $E$ and $B$ depending on the particular sky simulation.}
  
  \label{table1}
  \begin{center}
    \begin{tabular}{|c|c|c|c|c|}\hline
      Method & ${f_\mathrm{knee}/f_\mathrm{spin}}$ & $T$ & $E$ & $B$ \\ \hline
      white noise         & 0       & 1            & 1            & 1              \\ \hline\hline
      {\bf destriped}    & {\bf 0.5}& {\bf 1.19}  & {\bf 1.05}  & {\bf 1.03}   \\ \hline
      zero-averaged       & 0.5     & 51.9        &  9.23       & 3.64          \\ \hline
      undestriped         & 0.5     & 6.98        &  7.04       & 15.2          \\ \hline\hline
      {\bf destriped}    & {\bf 1} & {\bf 1.24}  & {\bf 1.12}  & {\bf 1.19}    \\ \hline
      zero-averaged       & 1       & 52.2        &  9.91       & 3.95          \\ \hline
      undestriped         & 1       & 10.7        &  10.9       & 7.51          \\ \hline\hline
      {\bf destriped}    & {\bf 2} & {\bf 1.26}  & {\bf 1.32}  & {\bf 1.23}    \\ \hline
      zero-averaged       & 2       & 49.5        &  10.2       & 3.85          \\ \hline
      undestriped         & 2       & 6.41        &  9.97       & 8.18          \\ \hline\hline
      {\bf destriped}    & {\bf 5} & {\bf 1.35}  & {\bf 1.39}  & {\bf 1.38}    \\ \hline
      zero-averaged       & 5       & 49.8        & 10.4        & 3.99          \\ \hline
      undestriped         & 5       & 11.3        & 8.24        & 12.4          \\ \hline
    \end{tabular}
  \end{center}
\end{table}

\begin{figure}[ht]
  \begin{center}
      \epsfig{ file=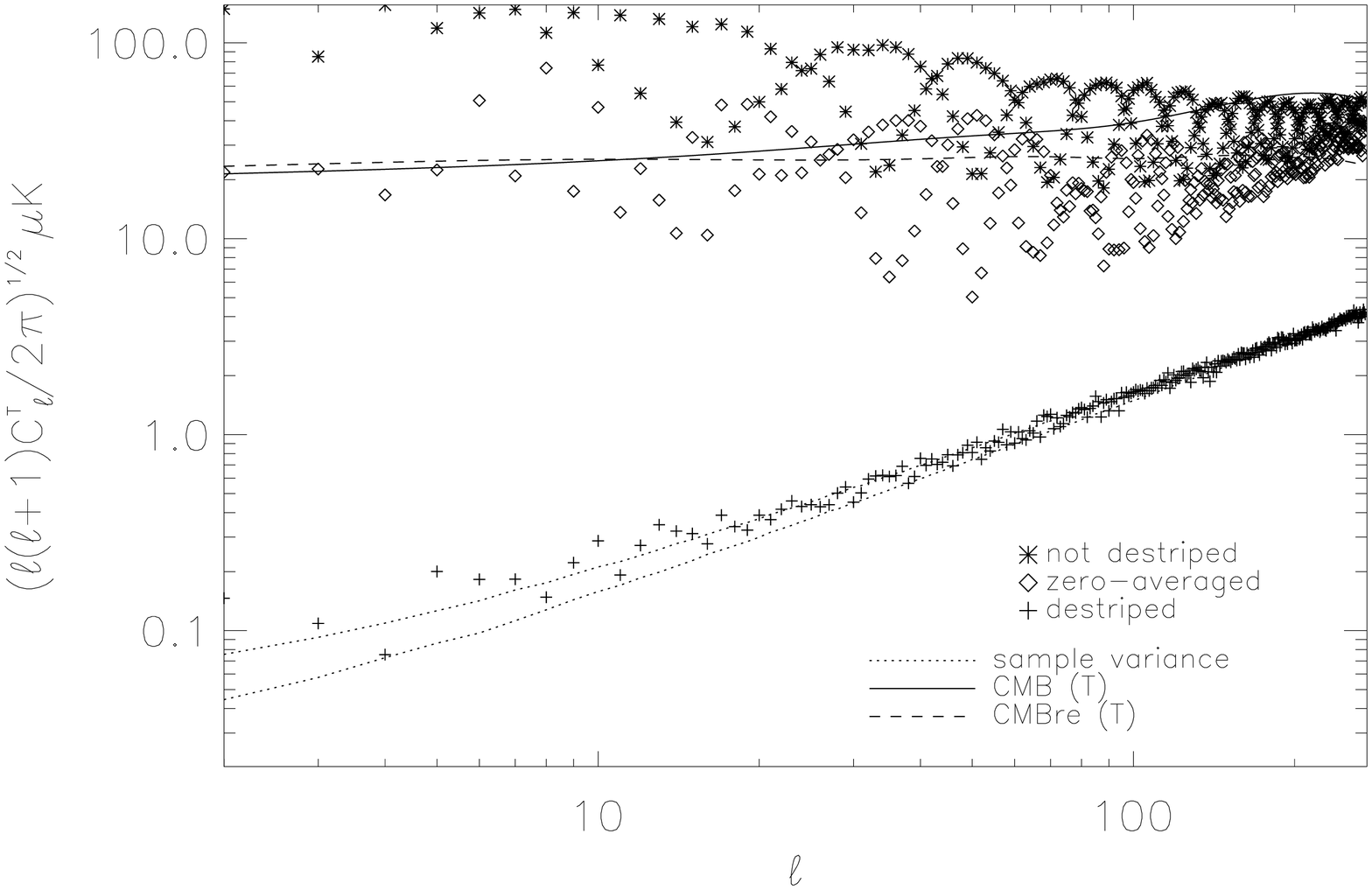,width=8.5cm}
    \end{center}
  \caption{Efficiency of destriping for the $T$ field with ${f_{\rm knee}/f_{\rm spin}=1}$.
    The sample variance associated to a pure white noise mission
    is plotted as the
    dotted lines.  The ``destriped spectrum'' is very close to the white noise
    spectrum (within the limits due to the sample variance). The zero-averaged
    and the ``not destriped'' spectra are a couple of orders of magnitude above.
    The solid line represents a standard CDM temperature spectrum and the dashed line represents
    a CDM temperature spectrum with reionization.
    \label{sptun}}
  \begin{center}
      \epsfig{ file=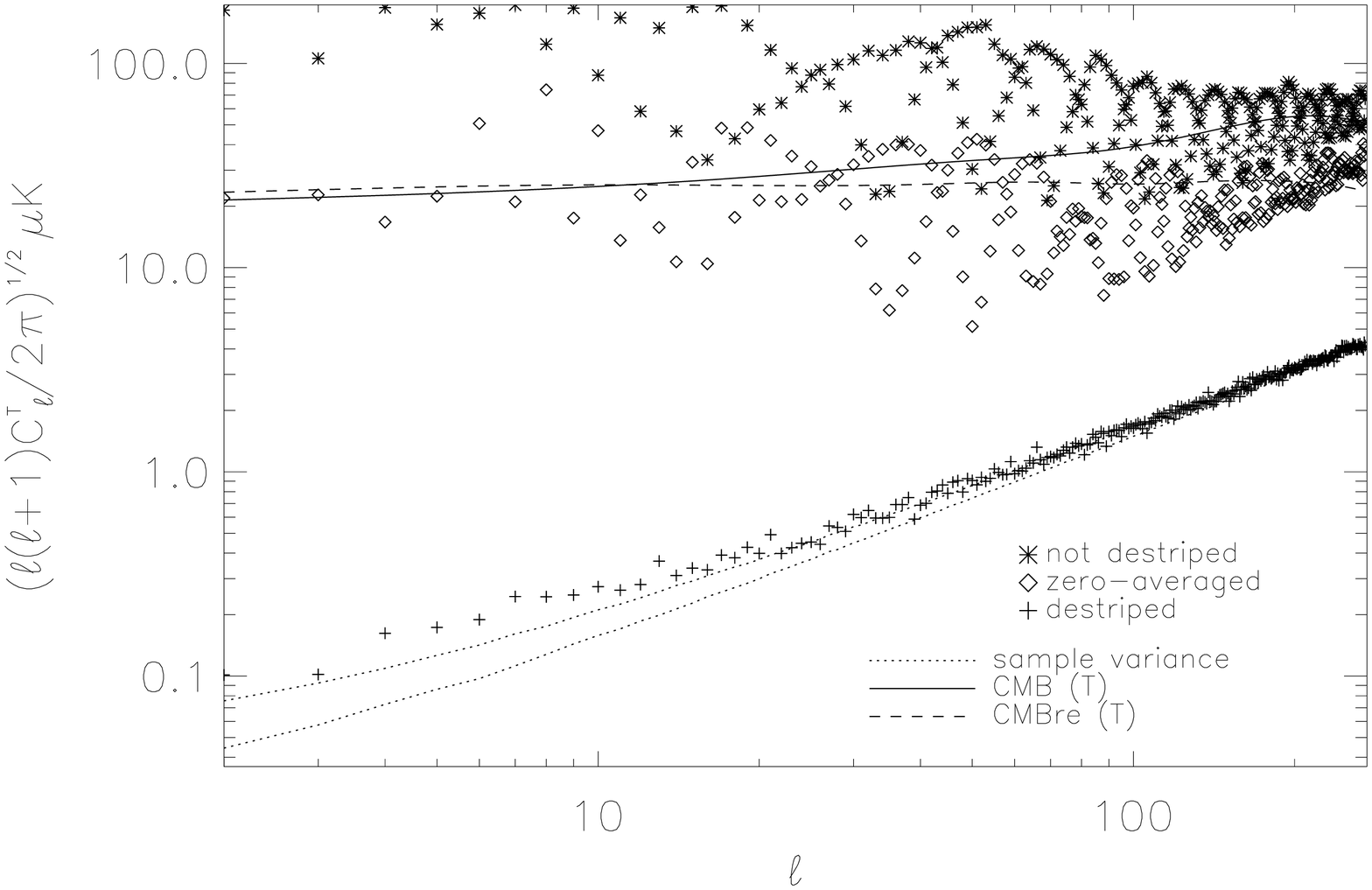,width=8.5cm}
    \end{center}
  \caption{Efficiency of destriping for the $T$ field with ${f_{\rm knee}/f_{\rm spin}=5}$. 
    The modelling of 
    low-frequency noise with an offset is no longer sufficient and
    the destriping leaves some power at low values of $\ell$. Nevertheless, 
    it remains a very good way to significantly reduce the effect of low-frequency noise.
    \label{sptcinq}}
\end{figure}
\begin{figure}[ht]
  \begin{center}
      \epsfig{ file=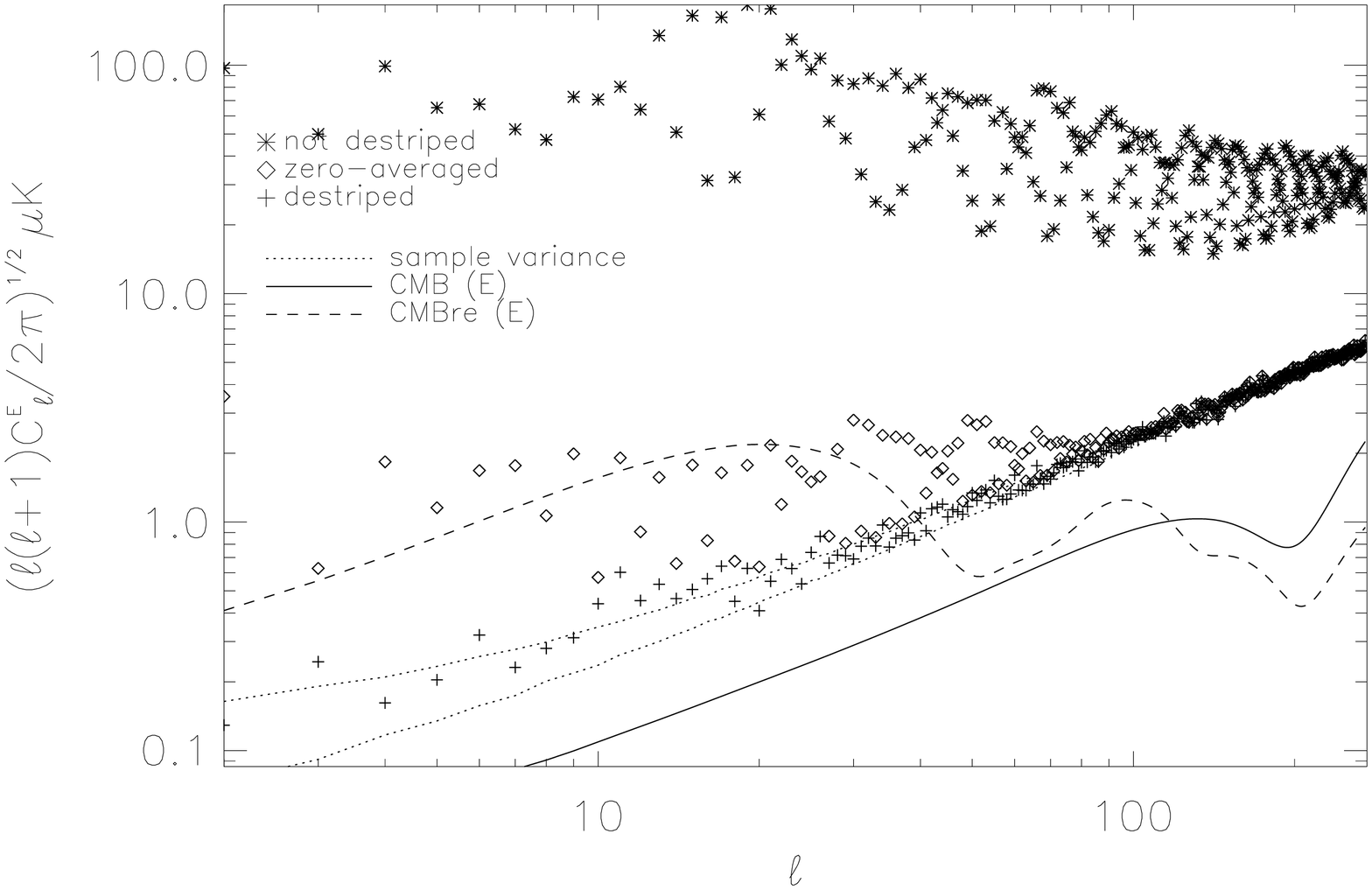,width=8.5cm}
    \end{center}
  \caption{Efficiency of destriping for the $E$ field for ${f_{\rm knee}/f_{\rm spin}=1}$. 
    The zero-averaged spectrum
    is not as bad as for $T$ but the residual striping we can see in Fig. \ref{sqmun}
    leads to some excess of power for low values of $\ell$ (up to $\ell\sim 100$). We do not see such effect in
    the destriped spectrum (and maps). The spectra for the $B$ fields are very similar.
    The solid line represents a standard CDM $E$ spectrum and the dashed line represents
    a CDM $E$ spectrum with reionization.      
    \label{speun}}
  \begin{center}
      \epsfig{ file=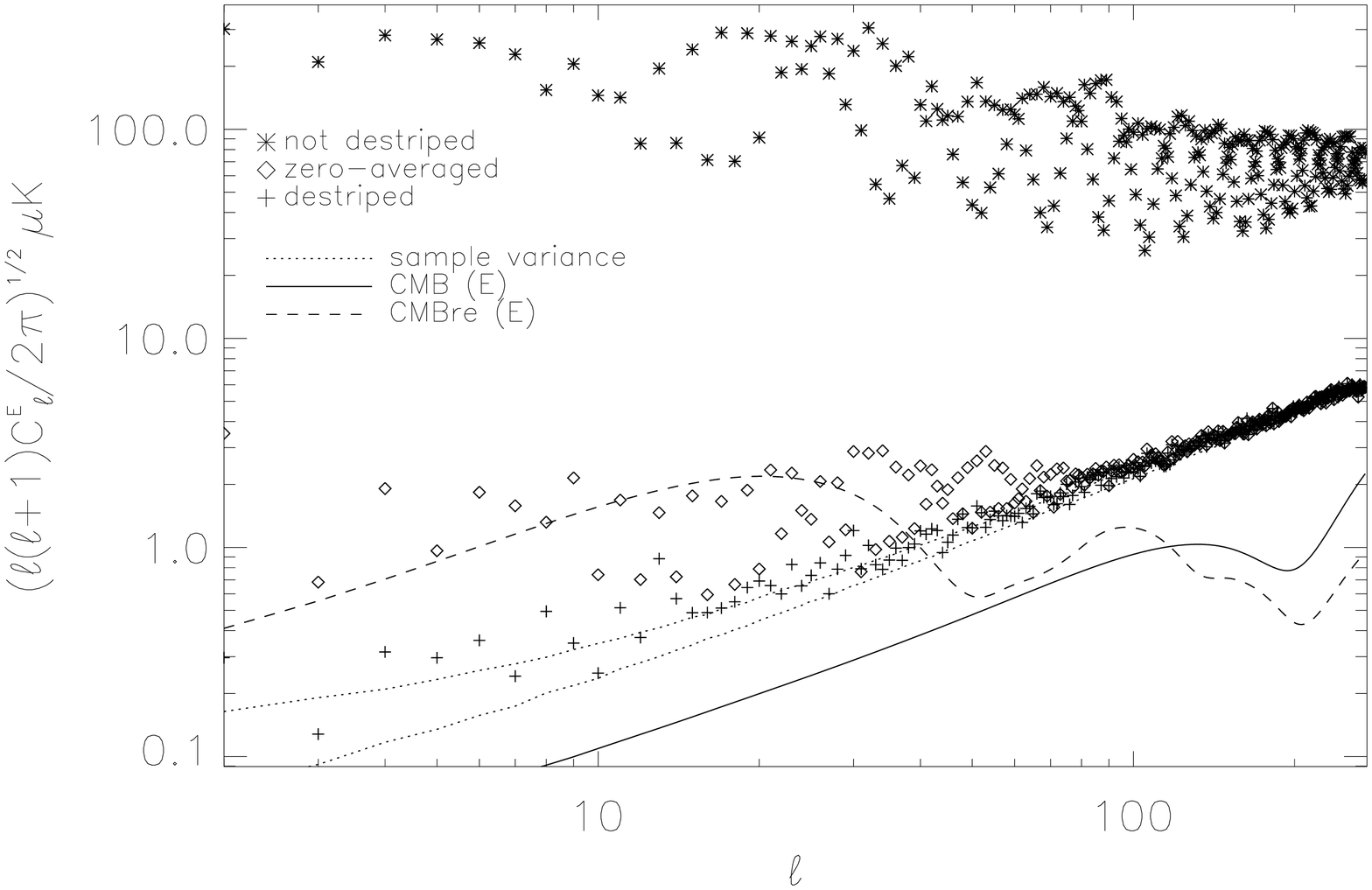,width=8.5cm}
    \end{center}
  \caption{Same as Fig. \ref{speun} but for ${f_{\rm knee}/f_{\rm spin}=5}$.
    The spectra for the $B$ fields are very similar.
    \label{specinq}}
\end{figure}
\begin{figure}[ht]
  \begin{center}
    \epsfig{ file=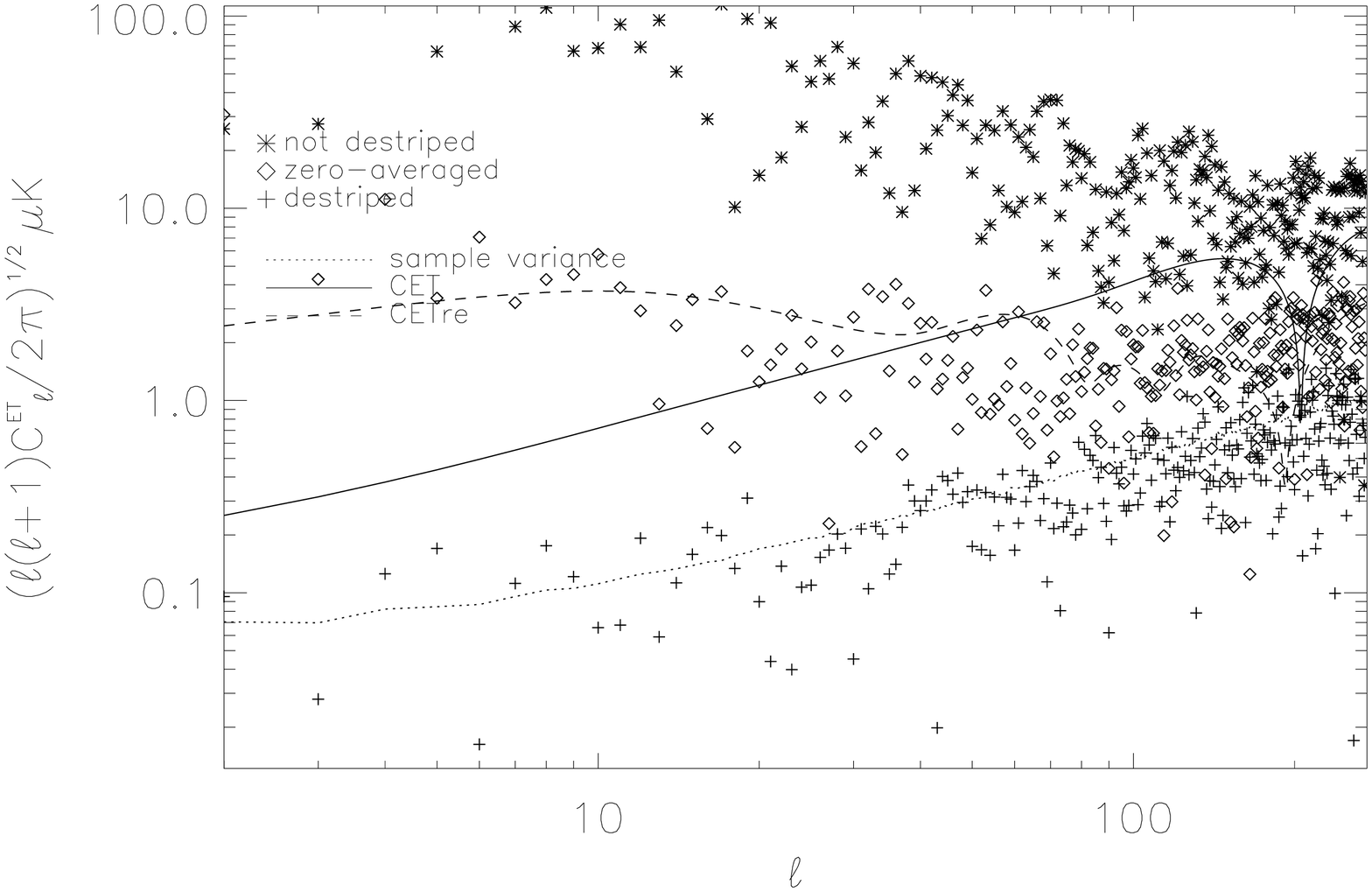,width=8.5cm}
  \end{center}
  \caption{Same as Fig. \ref{speun} for the $ET$-correlation for ${f_{\rm knee}/f_{\rm spin}=1}$. 
    \label{spetun}}
  \begin{center}
    \epsfig{ file=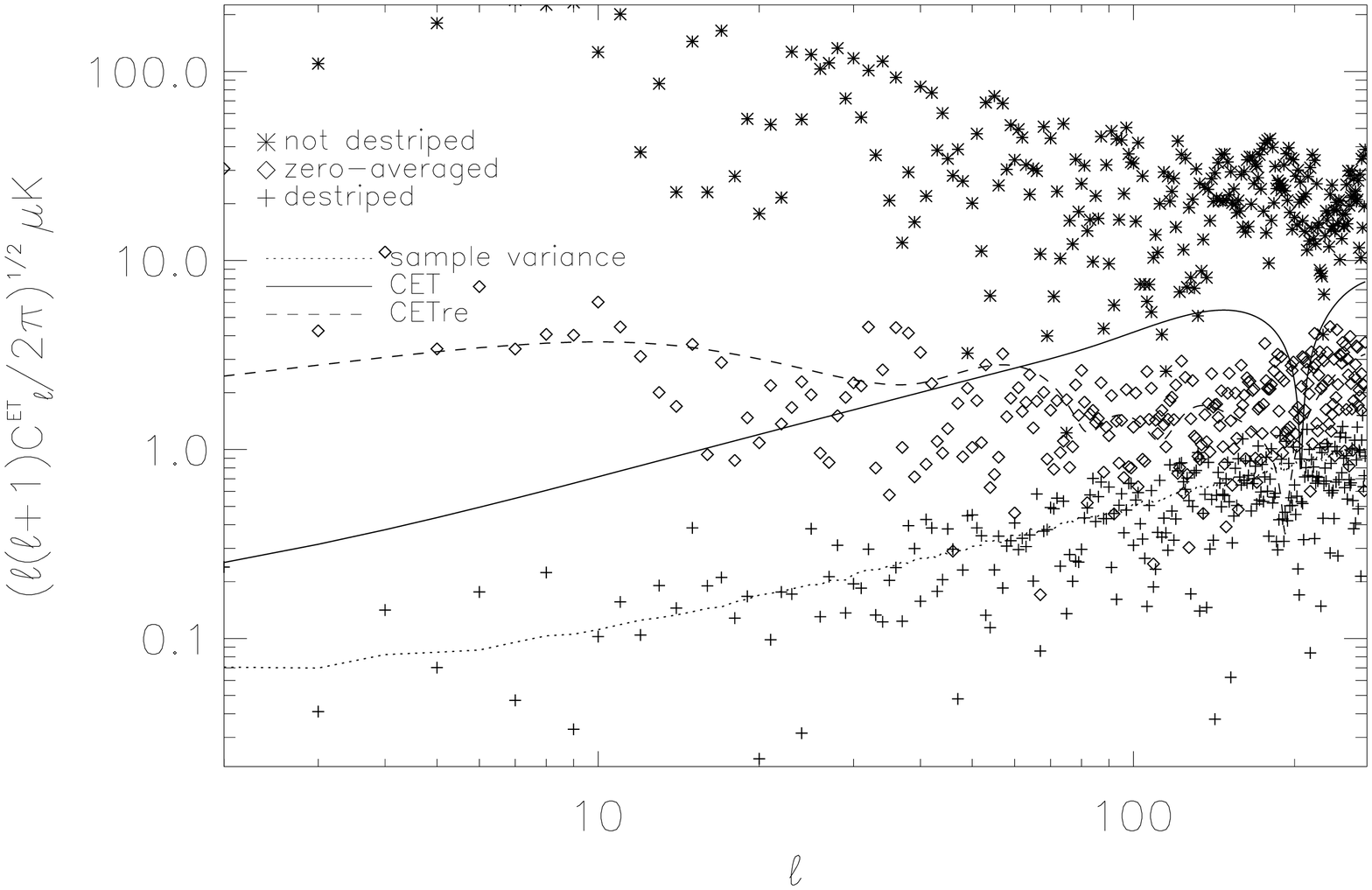,width=8.5cm}
  \end{center}
  \caption{Same as Fig. \ref{spetun} but for ${f_{\rm knee}/f_{\rm spin}=5}$.
    \label{spetcinq}}
\end{figure}

\section{Discussions and Conclusions}\label{resul}

\paragraph{Comparison with other methods.}

Although no other method has yet been developped specifically for destriping
polarized data, many methods exist for destriping unpolarized CMB data,
which could be adapted to polarized data as well.

We first comment on the classical method which consists in modelling the 
measurement as

\begin{equation}
m_{t} = A_{tp}T_{p} + n_{t}
\end{equation}

where $A$ is the so-called "pointing matrix", $T$ a vector of temperatures in
pixels of the sky, $m_{t}$ the data and $n_{t}$ the noise. The problem is
solved by inversion, yielding an estimator of the signal:

\begin{equation}
\tilde{T_{p}} = [A^t N^{-1} A]^{-1} A^t N^{-1} \, m,
\end{equation}
where $N=\langle nn^t \rangle$ is the noise correlation matrix and $A^t$ is the transposed matrix of $A$.

This method can be extended straightforwardly to polarized measurements,
at the price of extending by a factor of $3\times h$ the size of the matrix $A_{tp}$, by 3 that of
vector $T_{p}$ (replaced by $(I_{p},Q_{p},U_{p})$), and by $h$ that of the data
stream (remember that $h$ is the number of polarimeters). 
The implementation of this formally simple solution may turn into
a formidable problem when megapixel maps are to be produced. Numerical
methods have been proposed by a variety of authors 
\cite{wright96b,tegmark97}, that
use properties of the noise correlation matrix (symmetry,
band-diagonality) and of the pointing matrix (sparseness).
Such methods, however, rely critically on the assumption that the noise is
a Gaussian, stationary random process, which has been a reasonable
assumption for CMB mission as COBE where the largest part of the
uncertainty comes from detector noise, but is probably not so for sensitive
missions such as Planck. Our method requires only inverting a ${3n\times 3n}$ matrix
where $n$ is the number of circles involved, and does not assume
anything on the statistical properties of low frequency drifts. It just assumes
a limit frequency (the knee frequency) above which the noise can be considered as a white
Gaussian random process.

Another interesting method is the one that has been used by Ganga in the
analysis of FIRS data \cite{ganga94th}, which is itself adapted from a
method developed originally by Cottingham~\cite*{cottingham87}. In that method,
coefficients for splines fitting the low temperature drifts are obtained
by minimising the dispersion of measurements on the pixels of the map. Such
a method, very similar in spirit to ours, could be adapted to
polarization. Splines are natural candidates to replace our offsets in
refined implementations of our algorithm.

Here, we have assumed that the averaged noise can be modeled as circle
offsets plus white noise (Eq. \ref{eq:offsetsmodel}), i.e. that the noise between
different measurements from the same bolometer is uncorrelated after
removal of the offset.  This allowed us to simplify the $\chi^2$ to
that shown in Eq. \eqref{eq:chi2}.  In reality, the circular offsets do not
completely remove the low frequency noise and there does remain some
correlation between the measurements.  The amount of correlation is
directly related to the value of ${f_\mathrm{knee}/f_\mathrm{spin}}$; the smaller
${f_\mathrm{knee}/f_\mathrm{spin}}$ the smaller the remaining correlation.  
Figs. 13-18
already contain the errors induced from the fact that we did not
include these correlations in the covariance matrix, and thus
demonstrate that the effect is small for ${f_\mathrm{knee}/f_\mathrm{spin}\sim 1}$.

\paragraph{Conclusion.} The destriping as implemented in this paper 
removes low frequency drifts up to the white noise level provided that ${f_\mathrm{knee}/f_\mathrm{spin}\le 1}$.
For larger ${f_\mathrm{knee}}$, the simple offset model for the averaged noise could be
replaced with a more accurate higher-order model that destripes to
better precision provided the scan strategy allows to do so, as discussed in Delabrouille~et~al~\cite*{delabrouille98e}.
We are currently working on improving our algorithm to account for
these effects. However, despite the shortcomings of our model, it
still appears to be robust for small ${f_\mathrm{knee}}$ and can serve as a first order
analysis tool for real missions.
In particular, our technique cannot only be used for the Planck HFI
and LFI, but can also be adopted for other CMB missions with circular
scanning strategies, such as COSMOSOMAS for instance \cite{rebolo98}.

\begin{acknowledgement}
We would like to acknowledge our referee's very useful suggestions.
\end{acknowledgement}

\appendix
\section{Destriping of unpolarized data}\label{appendixa}

We give here the formul\ae\ for the simpler case of destriping
temperature measurements with bolometers. The assumptions are the same
as in the polarized case and we adopt the same notation for the common
quantities. Instead of polarimeters, we have $h$ bolometers.
Since the measurement is no longer dependent on the orientation of the
bolometer, the model of the measurement is given by:
\begin{equation}
\!\!\!\!\!\!\!\!\! \boldsymbol{M}_{i,j,\delta}=\boldsymbol{O}_i+I_{i,j,\delta}~\boldsymbol{u} \mathrm{~~with~~} 
\boldsymbol{u}=\left( 
\begin{array}{c}
1 \\
\vdots \\
1 
\end{array}
\right)
\end{equation}
where ${\boldsymbol{M}_{i,j,\delta}}$ is the $h-$vector made of measurements by 
the $h$ bolometers, ${\boldsymbol{O}_i}$ is an $h-$vector 
containing the offsets for the i'th circle and ${I_{i,j,\delta}}$ 
is the temperature in the direction of the intersection point labeled by 
$\{i,j,\delta\}$. $\boldsymbol{u}$ is an $h$-vector,
corresponding to the $\boldsymbol{\mathcal{A}}$ matrix of the polarized case.
The $\chi^2$ can be written as:
\begin{multline}
\!\!\!\!\!\!\!  \chi^2 = \sum_{i,j\in {\mathcal{I}(i)},\delta=
  \pm 1} \,\left(\boldsymbol{M}_{i,j,\delta} - \boldsymbol{O}_i -
    I_{i,j,\delta} \; \boldsymbol{u} \right)^T \times\\
  {\boldsymbol{N}_i}^{-1}\left(\boldsymbol{M}_{i,j,\delta} - \boldsymbol{O}_i -
    I_{i,j,\delta} \; \boldsymbol{u} \right).
\end{multline}
In this case, the physical quantity uniquely defined at an intersection point is 
the temperature of the sky at this point. The constraint used here for removing 
low-frequency noise is then:
\begin{equation}
\!\!\!\!\!\!\!\!\!  I_{i,j,\delta} = I_{j,i,-\delta}.\label{redundtemp}
\end{equation}
Given this relation, the minimization of the $\chi^2$ with respect to 
$\boldsymbol{O}_i$ and $I_{i,j,\delta}$ leads to the linear system:
\begin{multline}
\!\!\!\!\!\!  \sum_{j \in \mathcal{I}(i)} \frac{x_j}{x_i+x_j} \left( \Delta_i-\Delta_j \right) = \\ 
  \frac{1}{2} \sum_{j\in \mathcal{I}(i),\delta=\pm 1} \frac{x_j}{x_i+x_j} 
  \left( \mathscr{I}_{i,j,\delta} - \mathscr{I}_{j,i,-\delta} \right)
\end{multline}
where 
\begin{equation}
\!\!\!\!\!\!\!\!\!  x_i=\boldsymbol{u^T} \, \boldsymbol{N}_i^{-1} \, \boldsymbol{u}
\end{equation}
corresponds to the $\boldsymbol{X}_i$ matrix of the polarized case,
\begin{equation}
\!\!\!\!\!\!\!\!\! \mathscr{I}_{i,j,\delta}=\frac{1}{x_i} \, \boldsymbol{u^T} \, 
\boldsymbol{N}_i^{-1} \, \boldsymbol{M}_{i,j,\delta}
\end{equation}
corresponds to the $\mathscr{S}_{i,j,\delta}$ local Stokes parameters  of the polarized case and the scalar
\begin{equation}
\!\!\!\!\!\!\!\!\! \Delta_i=\frac{1}{x_i} \, \boldsymbol{u^T} \, \boldsymbol{N}_i^{-1} \,\boldsymbol{O}_i
\end{equation}
corresponds to the 3-vector $\Delta_i$ of the polarized case.

Temperature offsets appear
through their differences $\Delta_i - \Delta_j$ so this linear system
is not invertible and we can use the same methods to invert the system
than in the polarized case.
The size of the matrix to invert is $n \times n$. \\
After evaluation of the offsets $\Delta_i$, one can recover the value
of temperature for any sample $k$ along circle $i$:
\begin{equation}
\!\!\!\!\!\!\!\!\! I_{i,k}=\mathscr{I}_{i,k}-\Delta_i.
\end{equation}

\section{Details of the simulations}\label{appendixb}
The results presented in this paper correspond to the ``Optimized Configuration'' involving
$3$ polarimeters and to an angular step of $18^\prime$: the angle between two consecutive
samples along a circle is $18^\prime$ and there is one circle every $18^\prime$. 

In this case, the number of circles is $n_\mathrm{c}=1200$ and the number of samples on
each circle is $n_\mathrm{s}=1195$. The spin axis has a sinusoidal motion around the ecliptic plane
with an amplitude of $8^\circ$ and with $8$ cycles during the
mission. The aperture angle of the circles is $85^\circ$.
The simulated noise has ${f_\mathrm{knee}=\eta f_\mathrm{spin}}$ with ${\eta=1 \mathrm{~and~} 5}$. 

The white noise variance is calculated
based on the expected sensitivity on $Q$ and $U$ ($3.7~\mu$K/K) of the
(arbitrarily selected) 143~GHz polarized channel of the {\sc Planck} mission.

All the maps (CMB, dipole, galaxy, simulation) are HEALPIX maps with 196\,608 pixels of $27.48^\prime$.
Only 12 pixels ($0.006\%$ of the map) are not seen by the mission: their values are set to the average of the map.
The signal maps have been smoothed by a gaussian beam with a FWHM set to ${2.5\times 18^\prime}$.

We have run other simulations with ``Optimized Configuration'' involving
$3$ or $4$ polarimeters, and with cycloidal, sinusoidal or anti-solar spin axis trajectories 
(see Bersanelli et~al \cite*{bersanellicosa} and the {\sc Planck} web
page\footnote{\tt{http://astro.estec.esa.nl/SA-general/Projects\\
    /Planck/report/report.html}} for
additionnal information about proposed
scanning strategies). 
The aperture angle of the circles have been taken in $[70^\circ,75^\circ,80^\circ,85^\circ,90^\circ]$. 
In all these cases, the results are similar for ${f_\mathrm{knee}/f_\mathrm{spin}\sim 1}$.
For more pessimistic noise cases, the choice of the scanning strategy may have a strong impact on the quality of the
final maps. A quantitative study of this point is deffered to a forthcoming publication.

\listoffigures

\end{document}